\documentclass[twocolumn,showpacs,amsmath,amssymb,aps,prl,superscriptaddress]{revtex4}
\usepackage{epsfig,psfrag,amsmath,amssymb,float}
\usepackage{color}
\input{epsf}

\newcommand{\bc}{\begin{center}}
\newcommand{\ec}{\end{center}}
\newcommand{\be}{\begin{equation}}
\newcommand{\ee}{\end{equation}}
\newcommand{\beqn}{\begin{eqnarray}}
\newcommand{\eeqn}{\end{eqnarray}}

\def\1.2{\frac{1}{2}}

\def\af{antiferromagnetic}
\def\H{{\mathcal{H}}}
\def\S{\mbox{\bf S}}
\begin{document}
\title{Random-exchange quantum Heisenberg antiferromagnets on the square 
lattice}
\author{Nicolas Laflorencie}
\affiliation{Department of Physics \& Astronomy, University of British 
Columbia, Vancouver, B.C., Canada, V6T 1Z1}
\author{Stefan Wessel}
\affiliation{Institut f\"ur Theoretische Physik III, Universit\"at
  Stuttgart, 70550 Stuttgart, Germany}
\author{Andreas L\"auchli}
\affiliation{Institut Romand de Recherche Num\'erique en Physique des 
 Mat\'eriaux, EPFL, 1015 Lausanne, Switzerland}
\author{Heiko Rieger}
\affiliation{Theoretische Physik; Universit\"at des Saarlandes, 
  66041 Saarbr\"ucken; Germany}
\date{\today}

\begin{abstract}
The ground state properties of random-exchange spin-1/2 Heisenberg
antiferromagnets on the square lattice are investigated using a
combination of quantum Monte Carlo simulations, exact numerical
diagonalizations, and spin wave calculations.  Whereas arbirarily weak
disorder has a drastic effect on 1d Heisenberg AFM, we find that in
two dimensions
the characteristics of the ground state like long-range order is
robust even against strong disorder.
While the antiferromagnetic order parameter and the spin stiffness are
exponentially reduced for singular exchange distributions, they vanish
only in the limit of infinite randomness.
\end{abstract}
\pacs{75.40.Mg, 75.10.Nr, 02.70.Ss, 75.30.Ds}
\maketitle

The spin-1/2 Heisenberg antiferromagnet (AFM) on the square lattice
has attracted a lot of interest in the last two
decades~\cite{pure-huse,pure-young,NLSM,SE,Manousakis91,ED,Beard96,pure-sandvik},
motivated in particular by the suggestion that the high-temperature
superconductivity in cuprates is related to the magnetic properties of
their CuO planes~\cite{CuO}. Its ground state (GS) is
antiferromagnetically ordered and the value for the staggered
magnetization $m_{\rm AF}=0.3070(3)$ is known to high
accuracy~\cite{pure-sandvik}.  Recently it was found that this
magnetic long range order (LRO) is robust against the introduction of
static non-magnetic impurities, as observed in Mg or Zn doped
La$_2$CuO$_4$~\cite{Vajk02}. In fact, using numerical simulations, it
was shown that LRO persists up to the percolation threshold for site
dilution~\cite{dil-kato,dil-sandvik}. Furthermore, the critical
exponents for the transition to the paramagnetic phase are those of
the classical percolation transition~\cite{dil-sandvik}.

The question then arises, how the GS properties change for other forms
of quenched disorder, in particular for generic bond disorder,
i.e. randomness in the strength of the antiferromagnetic exchange
coupling.  In the one-dimensional (1D) case, where the GS of the pure
system is critical, bond disorder is relevant~\cite{Doty92}, and an
infinitesimal amount of bond disorder drives the system into the
random singlet phase with unconventional scaling
properties~\cite{1d-fisher,1d-igloi,1d-laflo}. On the other hand, the
gapped spin-liquid GS of antiferromagnetic spin-1/2 Heisenberg ladders
is stable against the introduction of bond
disorder~\cite{ladder-weak,ladder-strong}. Indications for similar
stability were obtained recently also for two-dimensional (2D)
Heisenberg antiferromagnets using the strong disorder renormalization
group~\cite{high-d,two-d}, which is, however, reliable only for the
opposite scenario, i.e.\ when the GS is described by an infinite
randomness fixed point scenario.

Here, we employ exact numerical methods to
reliably estimate the stability of the GS of 2D
Heisenberg
AFMs against exchange-randomness.
In particular, we consider the 
bond-disordered spin-1/2 Heisenberg AFM
on the square lattice, defined by the Hamiltonian
\be
\label{eq:model}
 \H=\sum_{\langle i,j\rangle}J_{ij}\ \S_i\cdot\S_j,
\ee
where $\langle i,j\rangle$ are nearest neighbor bonds on a $L\times L$
square lattice ($N=L^2$), $\S_i$ are spin-1/2 operators, and the AFM
exchange couplings $J_{ij}\ge0$ are quenched random variables with
probability distribution ${\mathcal{P}}(J)$. In the following, we will
consider two different types of disorder distributions: (i) a
$W$-dependent flat, bounded bond distribution, where the bonds are
uniformly and symmetrically distributed around $1$ with width $W$:
\be
\label{DistDis}
{\mathcal{P}}(J)=\Theta(J-(1-W))\cdot\Theta((1+W)-J)\,/\,2W
\ee
and (ii) a singular distribution controlled by the parameter
$\delta$ and given by~\cite{note1}
\be
\label{eq:dist}
{\mathcal{P}}(J)=
J^{-1+\delta^{-1}}\delta^{-1}\cdot\Theta(J)\Theta(1-J)\;.
\ee
This second type of disorder distribution allows us to study the stability of 
the GS towards a singular proliferation of weak bonds in the lattice structure.

We analyze the GS properties of the model (\ref{eq:model}) 
using a combination of quantum Monte Carlo (QMC) simulations, exact
diagonalizations (ED) based on the Lanczos method, 
and spin wave theory (SW). In particular, we consider the stability of the 
staggered magnetization and the spin stiffness towards
bond disorder, using  disorder averaged quantities as obtained from several 
thousand disorder realizations from the distributions (2) and 
(3) with different parameters $W$ or $\delta$, respectively.

\begin{figure}
\bc
\includegraphics[width=\linewidth,clip]{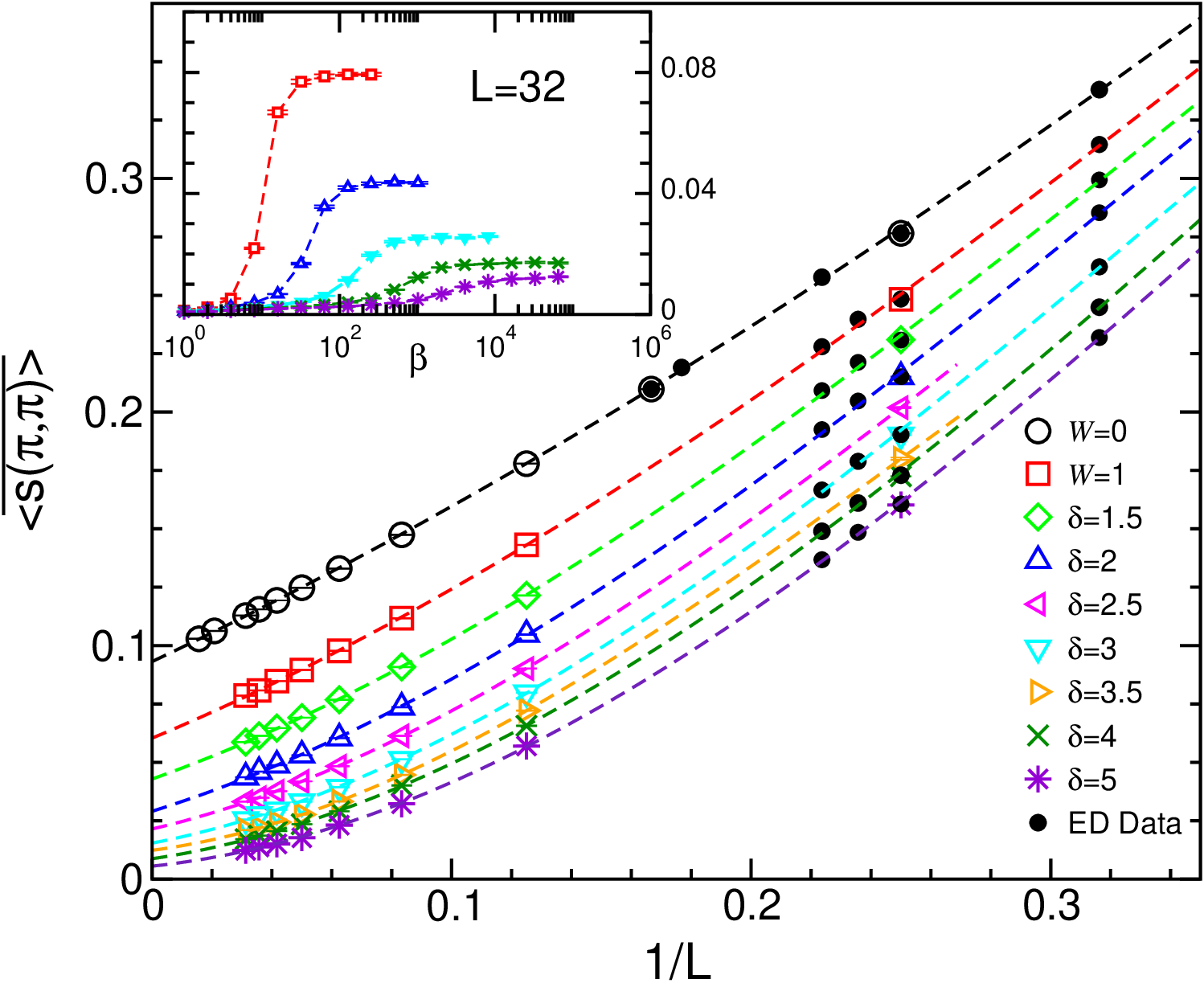}
\caption{
  (Color online) Disorder averaged value of the GS staggered structure
  factor ${\overline{\langle s(\pi,\pi)\rangle}}$ as a function of $1/L$. 
  Filled symbols are ED data and open symbols QMC data.
  The number of random samples varied between 1000 for the
  largest lattice and 10000 for the smaller lattices. 
  Dashed lines denote  third order
  polynomial fits of the finite size data. 
  The inset exhibits the convergence of $\langle s(\pi,\pi)\rangle$ using
  $\beta$-doubling scheme for 
  $\delta=1,~2,~3,~4,~5$ (top to bottom) and $L=32$.
  \label{fig:fss} 
}
\ec
\end{figure}

A N\'eel-ordered GS in the thermodynamic limit is signaled by a
non-vanishing infinite system size limit $m_{\rm{AF}}^2$ of the
staggered structure factor per site $s(\pi,\pi)$:
\be
s(\pi,\pi)=\frac{3}{N^2}\left\langle(\sum_{i=1}^{N}(-1)^i
  S_{i}^{z})^2\right\rangle\to m_{\rm{AF}}^2~~(N\to\infty).
\label{eq:spi}
\ee
The QMC simulations have been performed using the 
stochastic series expansion technique  with a directed loop update 
scheme~\cite{Sandvik02}.
For the clean (i.e. no disorder) case, we studied square lattices 
up to $L=64$. We found that a scaling of the inverse temperature
as $\beta=8\times L$ allows to  obtain
GS properties.
For the random case we needed to perform the QMC simulations at
much lower temperature, as correlations on weak bonds (i.e. small $J_{ij}$) 
still develop at very low temperatures.
To stimate GS quantities reliable, we
used the $\beta$-doubling scheme~\cite{dil-sandvik} which accelerates
the simulations considerably at low temperatures. We were able to
extract the GS properties of systems with up to $32\times 32$ lattice sites,
for values of the disorder-parameter up to $\delta=5$ by doubling
$\beta$ up to $2^{16}$ (see the inset of Fig.~\ref{fig:fss}). 
For the small lattices we have also carefully checked the convergence against
the ED results.
The number of random samples considered for disorder averaged
quantities varied between 1000 for the largest lattice and 10000 for
smaller lattices.

In Fig.~\ref{fig:fss} we show the disorder averaged GS values of
${\overline{\langle s(\pi,\pi) \rangle}}$, which we obtained from the
QMC and ED calculations as functions of $1/L$.  For the clean case
($\delta=0$) QMC data for lattices ranging from $L=4$ up to $L=64$
have been extrapolated to the thermodynamic limit using an
unconstrained third order polynomial fit in $1/L$. The obtained
staggered magnetization $m_{\rm{AF}}=0.3064(2)$, is in good agreement
with previous estimates~\cite{pure-sandvik,note}. Increasing the
randomness $\delta$, we found that large system sizes are required to
reliably extrapolate to the thermodynamic limit.  While in the clean
case and for weak disorder the finite size scaling is dominated by a
term linear in $1/L$, higher-order terms become important for larger
values of $\delta$.  This can be seen from Fig.~\ref{fig:fss}, where
for strong disorder and large systems sizes, the curvature of the
polynomial extrapolations become more pronounced, indeed requiring up
to third order polynomials in $1/L$.  We checked that the extrapolated
values did not change within the statistical error bars upon allowing
higher order fitting polynomials.

To further analyze the effects of the randomness on the quantum
fluctuations above the classical N\'eel state, we employ linear
spin-wave~\cite{sw_sw} theory. This appears justified given the
finite, although strongly reduced, staggered moment
obtained within the QMC simulations.  As translational symmetry is
broken in each realization for the disordered system, we adopt a
a real-space formulation of the spin-wave approach to finite
systems. In the following, we use the method of
Ref.~\cite{sw_quasicrystal}, which is based upon solving a
non-hermitian eigenvalue problem and subsequent orthogonalization to
obtain the bosonic eigenmodes of the quadratic part of the spin-wave
Hamiltonian. Similar approaches have been used recently to study
site-dilution effects in the Heisenberg model on the
square~\cite{sw_square} and honeycomb~\cite{sw_hexagonal}
lattice. 
For details on the
numerical scheme, we refer to Ref.~\cite{sw_quasicrystal}. 
\begin{figure}
\bc
  \includegraphics[width=0.9\linewidth,clip]{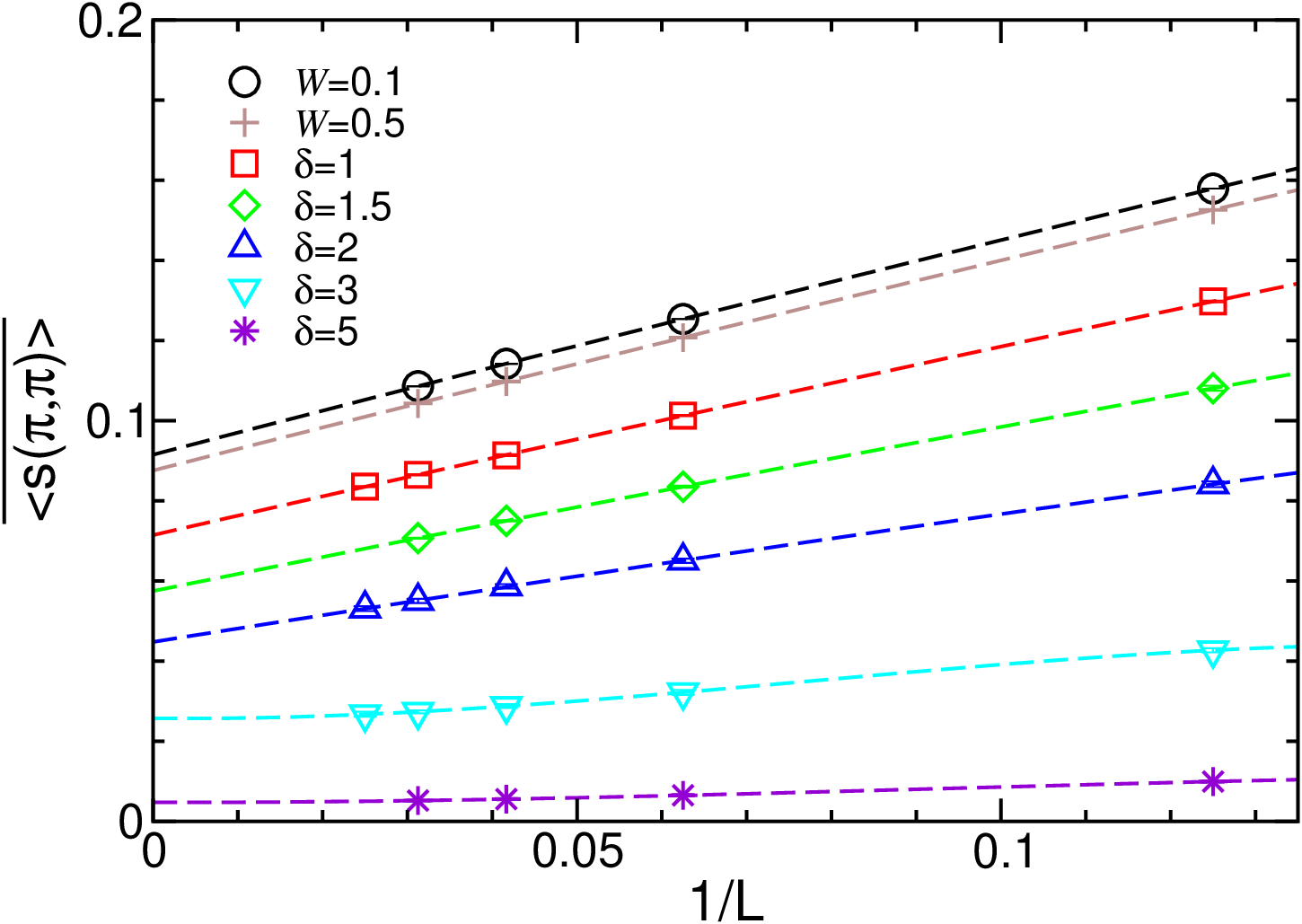} 
  \caption{(Color online) Results from
  spin wave theory for the disorder averaged value of the GS staggered
  structure factor ${\overline{\langle s(\pi,\pi)\rangle}}$ as a
  function of $1/L$. Dashed lines denote polynomial fits in $1/L$.
  \label{fig:sw} } \ec
\end{figure}

We perform the above procedure for typically 1000 disorder
realizations for system sizes $L=8, 16, 24, 32$ and $40$ for various
values of $\delta$.  The spin-wave estimate of ${\overline{\langle
s(\pi,\pi)\rangle}}$ for given values of $L$ and $\delta$ is finally
obtained by averaging the local staggered moments from the various
disorder realizations.  We found that for $\delta>1$ a small fraction
of sites attain negative values of the local staggered moment.  We
treat this artifact of the linear spin-wave theory, related to the
singular disorder distribution at $J=0$, by explicitly setting the
local staggered moments of the corresponding sites to zero before
performing the averaging procedure. Results of the SW calculation are
shown in Fig.~\ref{fig:sw} for various disorder strengths as functions
of $1/L$.  Extrapolations to the thermodynamic limit have been
performed using third order polynomial fits in $1/L$, where again 
high-order terms become important as the randomness increases.

\begin{figure}
\bc
  \includegraphics[width=0.9\linewidth,clip]{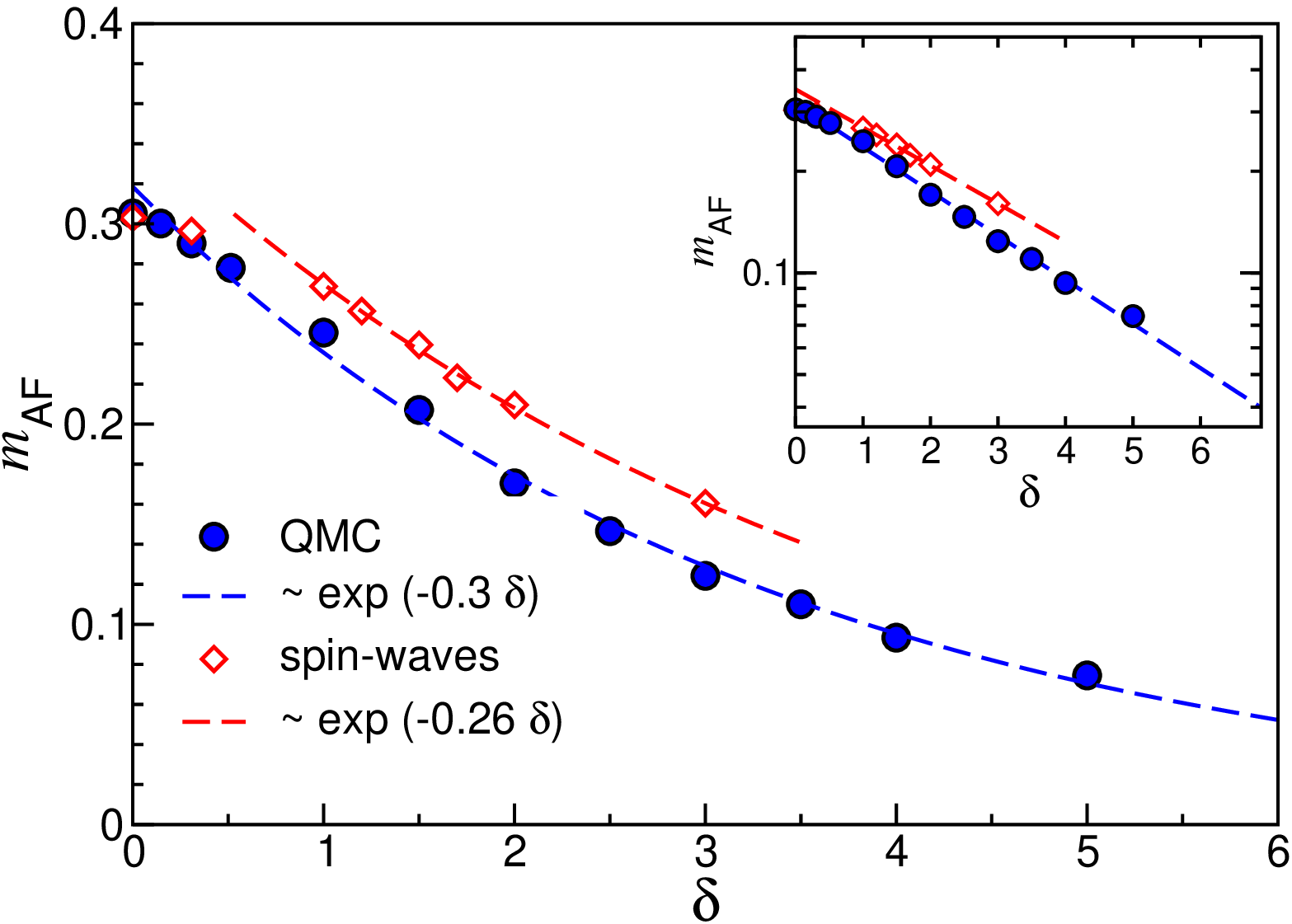} 
  \caption{(Color online) Disorder
  averaged \af~order parameter as a function of the disorder strength
  $\delta$. Dashed lines denote exponential fits of the data for
  $\delta> 1$, and the inset shows the same data on a semi-logarithmic
  plot. Note also that error bars are smaller than symbol sizes.
  \label{fig:mAF} }
\ec
\end{figure}

Based on our finite size scaling analysis, we now consider the
behavior of the staggered magnetization $m_{\rm{AF}}$ in the presence
of bond-randomness. We plot the values obtained from both the ED and
QMC and the SW calculations as functions of $\delta$ in
Fig.~\ref{fig:mAF}.

We find the N\'eel order is very robust against bounded bond
randomness. While the SW approach slightly overestimates the order
parameter (as expected), the form of the $\delta$-dependence of
$m_{\rm{AF}}$ calculated within SW theory agrees well with the ED and
QMC results. Using the distribution of Eq.~(3), the order parameter
shows an exponential decay (c.f. the inset Fig.~\ref{fig:mAF}) of the
form
\be
m_{\rm{AF}}\sim \exp(-c_m\delta),
\ee
with $c_m=0.301(5)$ obtained from the ED and QMC calculations, and
$c_m= 0.260(5)$ within SW theory. This suggests that the robust LRO
vanishes only in the limit of infinite randomness, i.e. for
$\delta\to\infty$.

In addition to the staggered magnetization, the ordered nature of the
Heisenberg antiferromagnet is accompanied by a finite value of the
spin stiffness, defined as the second derivative of the GS energy with
respect to a twist angle $\Phi$ introduced at the boundary along a
direction perpendicular to the order parameter:
\be
\rho_{s}=\frac{3}{2}\frac{\partial^2 E_{0}}{\partial \Phi^2}\Big|_{\Phi=0}.
\ee
In contrast to $m_{\rm{AF}}$, the spin stiffness is a dimensionful
quantity, and in the clean case, where $\rho_s=0.175(2)J$, scales
proportional to the exchange constant $J$.  A naive expectation would
suggest that in the disordered case, $\rho_s$ scales with the averaged
value of the exchange, $\overline{ J } \sim \frac{1}{\delta+1}$,
leading to an algebraic decay for large values of $\delta$.  In the
following, we show that instead $\rho_s$ shows an exponential decay,
similar to the staggered magnetization, $m_{\rm{AF}}$. From the QMC
simulations, $\rho_s$ can be evaluated for each disorder realization
by measuring the world-line winding number
fluctuation~\cite{pure-sandvik}, and finally performing a disorder
averaging.  In a clean system, a twist $\Phi$ enforced at the
boundaries will be equally distributed among the bonds along the
corresponding direction, resulting in a homogeneous twist in the order
parameter of $\Phi/L$ per bond.  In contrast, for a bond-disordered
system, larger twists occur for weaker bonds~\cite{Paramekanti98},
resulting in an inhomogeneous distribution of the local twist angle.
The winding number however, being a global quantity, provides an
estimate for the global rigidity of the system under such a twist in
the boundary conditions.

The spin stiffness can be expressed as
$\rho_s=-\frac{e_0}{2}-\frac{3\Lambda_{s}^{0}}{2}$, in terms of GS
energy per site without twist, $e_0$, and $\Lambda_{s}^{0}$, the
$\omega=0$ current-current correlation function
$\Lambda_s(\omega)=\frac{1}{L^2}\int_{0}^{\beta}d\tau
{\rm{e}}^{-i\omega\tau}\langle
j_s(\tau)j_s(0)\rangle$~\cite{Schulz95,pure-sandvik}.  According to
spin-wave calculations~\cite{Schulz95,Huse88}, in the clean case
$\Lambda_{s}^{0}(L)$ scales like
$\Lambda_{s}^{0}(L)=\Lambda_{s}^{0}+\frac{a_1}{L}+\frac{a_2}{L^2}+\hdots$,
whereas the GS energy scales as $e_0(L)-e_0\sim
\frac{1}{L^3}$~\cite{FSEnergy,pure-sandvik}.
\begin{figure}
\bc
\includegraphics[width=\linewidth]{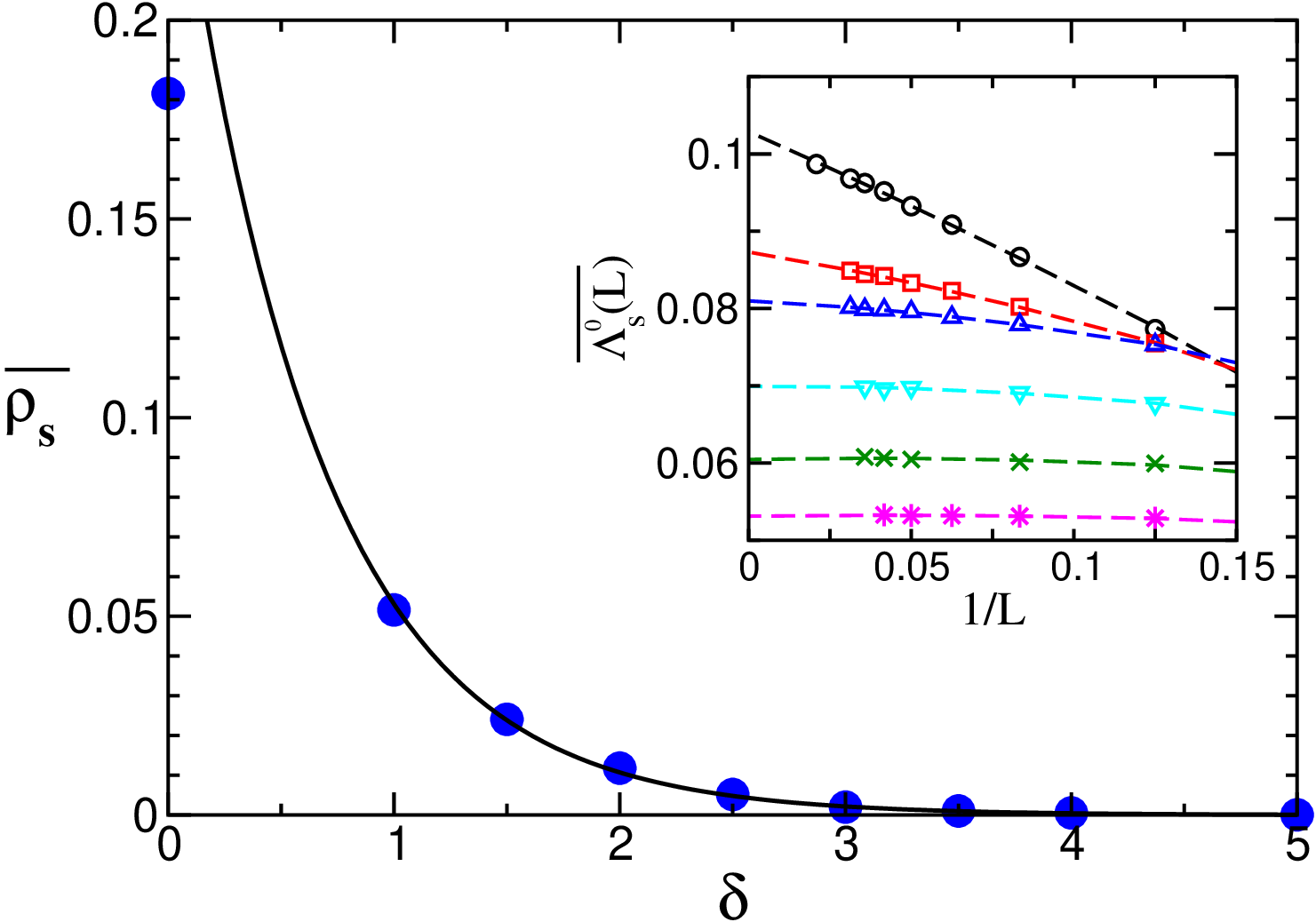}
\caption{(Color online) Disorder averaged values of the spin 
  stiffness ${\overline{\rho_s}}$ obtained
  from QMC simulations on systems with up to $L=32$ and carefull
  finite size scaling to the thermodynamic limit (see text). The solid
  line denotes an exponential fit Eq. (7). Inset: Finite size scaling
  of the disorder average $\omega=0$ current correlator
  ${\overline{\Lambda_{s}^{0}(L)}}=-{\overline{e_0(L)}}/3-
  2{\overline{\rho_s(L)}}/3$ plotted vs $1/L$ for
  $\delta=0,~1,~2,~3,~4,~5$ (different symbols from top to
  bottom). Dashed lines are quadratic fits. Error bars are smaller
  than symbol sizes.}  \label{fig:stiff} \ec
\end{figure}
We find that the $1/L^3$ scaling of $e_0(L)$ also holds for the
disordered averaged values. To reliably obtain the disorder average
spin stiffness ${\overline{\rho_s}}$ in the thermodynamic limit, we
subtracted the energy contribution to the stiffness and performed
polynomial fits in $1/L$ for the disorder averaged values of
${\overline{\Lambda_{s}^{0}(L)}}$, as shown in the inset of
Fig.~\ref{fig:stiff}. Interestingly, for increasing disorder, finite
size effects get strongly reduced which implies that the thermodynamic
limit value for ${\overline{\rho_S}}$ is, within error bars, reached
for smaller systems as disorder increases.  Since more QMC steps are
needed for each random sample to compute the fluctuations in the
winding numbers, than for the structure factor, the disorder average
was performed over slightly less random samples. Typically $1000$
samples were used for the smaller and $200$ for the largest lattice
with $L=32$.  Our results for ${\overline{\rho_s}}$ as a function of
$\delta$ are shown in Fig.~\ref{fig:stiff}. We find that the spin
stiffness remains finite upon increasing the disorder, showing an
exponential reduction with $\delta$, similar to the AF order
parameter.  In fact, a decay
\be
\rho_s(\delta)\sim \exp(-c_\rho\delta),
\label{eq:stiff}
\ee
with $c_\rho=1.60(2) $, fits the QMC data for $\delta>1$. 

In conclusion we studied the ground state properties of the
bond-disordered spin-1/2 Heisenberg model on the square lattice, and
found the antiferromagnetic order to be rather robust against
bond-randomness: Both the staggered magnetization and the spin
stiffness get reduced exponentially with $\delta$, but vanish only in
the limit of infinite randomness. These observations suggest a
reduced relevance of disorder effects on quantum Heisenberg models in
dimensions $D>1$. Indeed, and in contrast to the Ising model in a
transverse field, where disorder is relevant also for
$D>1$~\cite{Pich94,IRFP}, the infinite randomness fixed point is
unstable in XY and Heisenberg AFMs for $D\ge 2$~\cite{high-d}. This
result of a renormalization group study is confirmed here for the
Heisenberg case on the square lattice. It would be interesting to
achieve an analytical understanding of the exponential decay we
obtained for both the order parameter and the spin stiffness.  In an
earlier study, performed for a similar random-exchange AFM, a finite
critical disorder strength was found to destroy N\'eel order and to
drive the system into a paramagnetic phase~\cite{Vekic95}. Based on
simulations using significantly larger system sizes at essentially
zero temperature, we conclude on the absence of any such finite
critical disorder strength. We furthermore expect a similar robustness
of N\'eel-ordered GSs of quantum AFMs on other bipartite lattices,
such as the honeycomb lattice. In contrast, for the case of a
triangular lattice the interplay between randomness and frustration is
expected to destroy the ordered GS of the pure case~\cite{Claire}, and
drive the system into a spin-glass phase for strong
bond-disorder~\cite{high-d}.  Finally, the question arises, how the
excitation spectrum and finite temperature properties are affected by
bond-disorder.  Preliminary results \cite{unpub} indicate that upon
increasing the disorder strength, the character of the low-energy
excitations changes from extended spin waves to localized spin-cluster
flips.

We acknowledge valuable discussion with A. Sandvik, Y.-C. Lin and F.\
Igl\'oi.  The research of N.L. was supported by NSERC of Canada, and
the numerical simulations were carried out on the WestGrid network and
at NIC J\"ulich.
\vskip-0.5cm


\begin{thebibliography}{}
\bibitem{note1} The disorder
  strength $\delta$ can be defined as the variance of a random variable $\ln J$
  which in case of a flat distribution is related to $W$ according
  to~$\delta^2={1-\frac{1-W^2}{4W^2}\Bigl[\ln\Bigl(\frac{1+W}{1-W}\Bigr)\Bigr]^2}$.
\bibitem{note} Note the value obtained in Ref.~\cite{pure-sandvik}
  using an unconstrained finite size scaling fit is 0.3062(6) and
  0.3070(3) with a constrained fit.
\end{thebibliography}
\end{document}